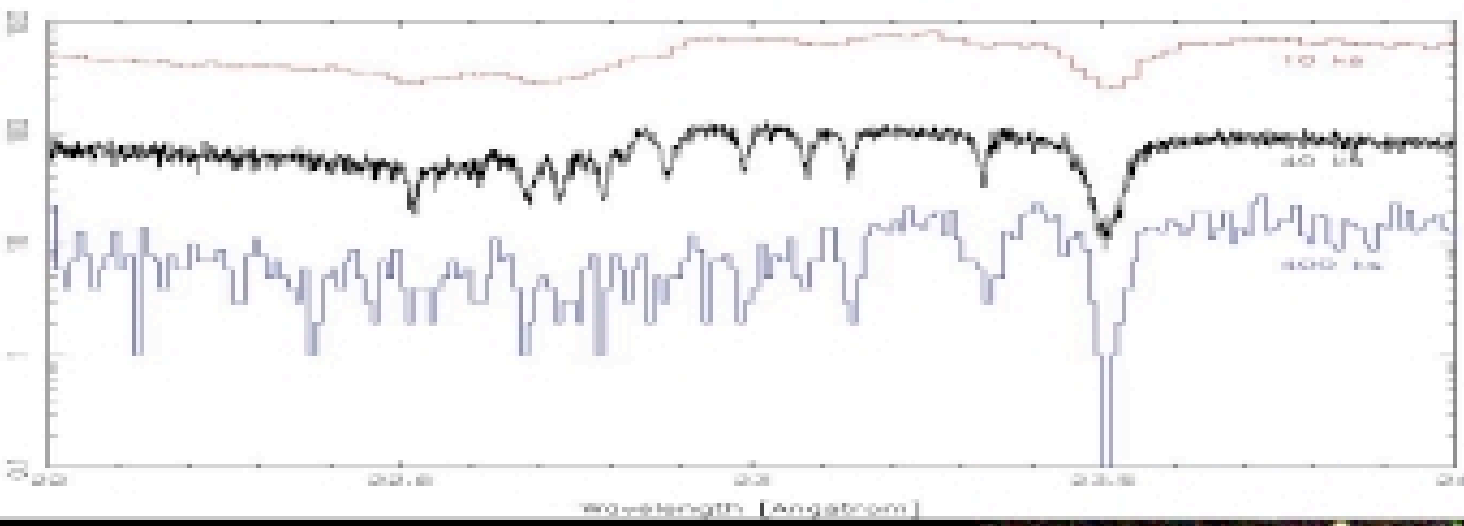

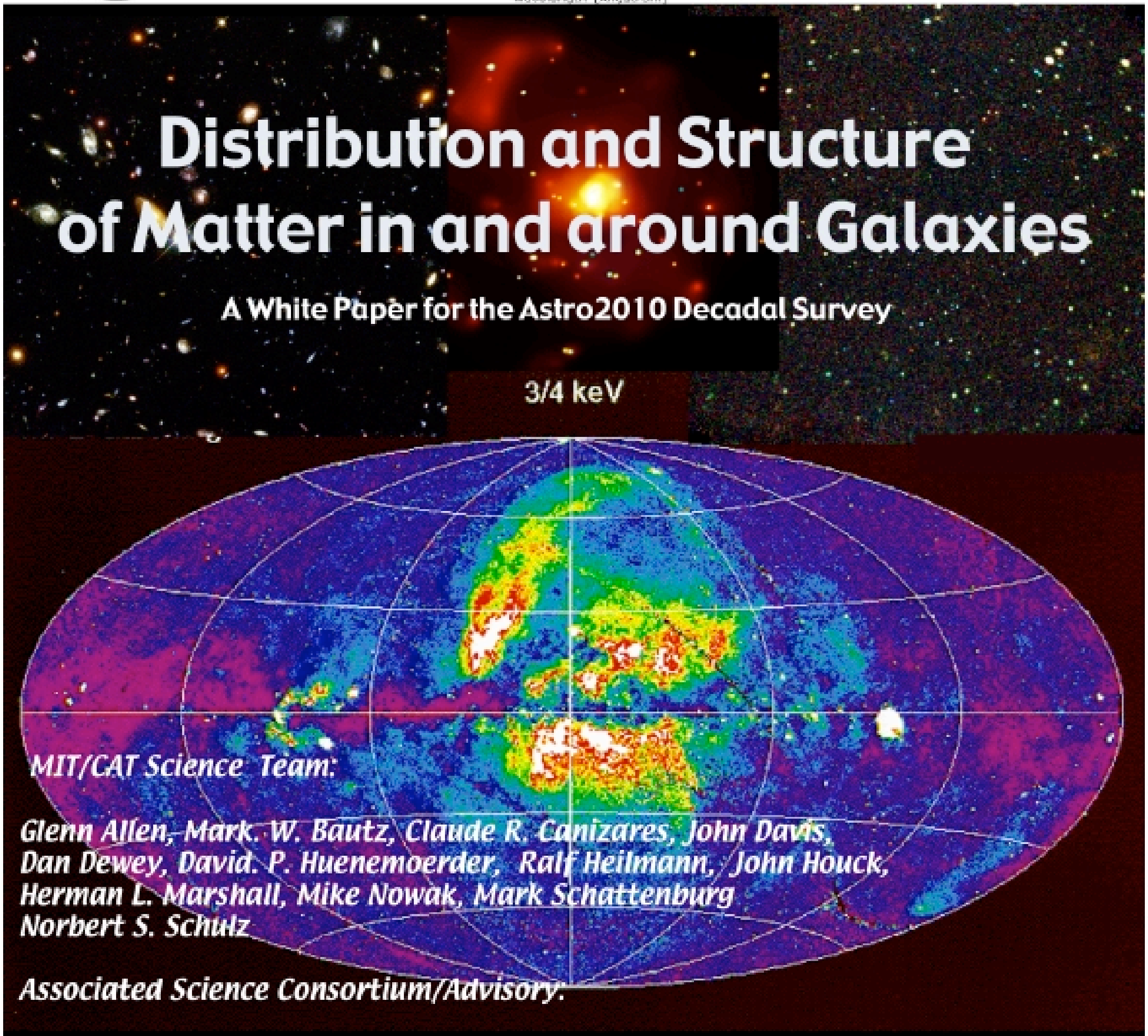

# Distribution and Structure of Matter in and around Galaxies

## A White Paper for the Astro2010 Decadal Survey

3/4 keV

*MIT/CAT Science Team:*

*Glenn Allen, Mark. W. Bautz, Claude R. Canizares, John Davis, Dan Dewey, David. P. Huenemoerder, Ralf Heilmann, John Houck, Herman L. Marshall, Mike Nowak, Mark Schattenburg Norbert S. Schulz*

*Associated Science Consortium/Advisory:*

*Joel Bregman (Michigan), Maria Diaz Trigo (ESA), Taotao Fang (UCIrvine) , Marc Gagne (WCU), Tim Kallmann (GSFC), Maurice Leutenegger (GSFC), Julia Lee (Harvard), Jon Miller (Michigan ), Koji Mukai (GSFC), Frits Paerels (Columbia), Andy Pollock (ESA), Andy Rasmussen (SLAC), John Raymond (CfA), Randall Smith (CfA), Yangsen Yao (CASA)*

*Comtact author: Norbert S. Schulz (MIT), nss@space.mit.edu*

***Abstract***

*Understanding the origins and distribution of matter in the Universe is one of the most important quests in physics and astronomy. Themes range from astro-particle physics to chemical evolution in the Galaxy to cosmic nucleosynthesis and chemistry in an anticipation of a full account of matter in the Universe. Studies of chemical evolution in the early Universe will answer questions about when and where the majority of metals were formed, how they spread and why they appar today as they are. The evolution of matter in our Universe cannot be characterized as a simple path of development. In fact the state of matter today tells us that mass and matter is under constant reformation through on-going star formation, nucleosynthesis and mass loss on stellar and galactic scales. X-ray absorption studies have evolved in recent years into powerful means to probe the various phases of interstellar and intergalactic media. Future observatories such as IXO and Gen-X will provide vast new opportunities to study structure and distribution of matter with high resolution X-ray spectra. Specifically the capabilities of the soft energy gratings with a resolution of R=3000 onboard IXO will provide ground breaking determinations of element abundance, ionization structure, and dispersion velocities of the interstellar and intergalactic media of our Galaxy and the Local Group.*

## 1. Distribution of Matter in the Universe

Concepts to understand the texture of matter date back to ancient times, it was during the middle of the 20th century with the formulation of the Standard Models of particle physics and cosmology where the physical groundwork for the treatment of matter in the Universe was laid. Today this involves atoms, nucleons, quarks and leptions, as well as mediators of fundamental forces, i.e. photons, gluons, and bosons. The chemical evolution of the Universe embraces aspects that reach deep into modern astrophysics and cosmology. We want to know, starting from observed distributions of element abundances here on Earth and our Solar System, over the study of stellar populations, the interstellar medium in the Milky Way and other galaxies, how present and past dated matter is affected by various levels and types of nucleosynthesis and stellar evolution and possibly trace this evolution back to the Big Bang. How daunting this task really is may be seen in Figure 1 (left). Three major constituents are highlighted within the gray box, which represent the study of primordial star formation including periods of super-massive black hole formation, the embedded evolution of the intergalactic medium (IGM), and finally the status and evolution of the interstellar medium (ISM) in and around galaxies. The latter is the focus of this paper.

Little do we know today about the relative abundance, ionization state, and distribution of light elements in the Z range from Carbon to Magnesium

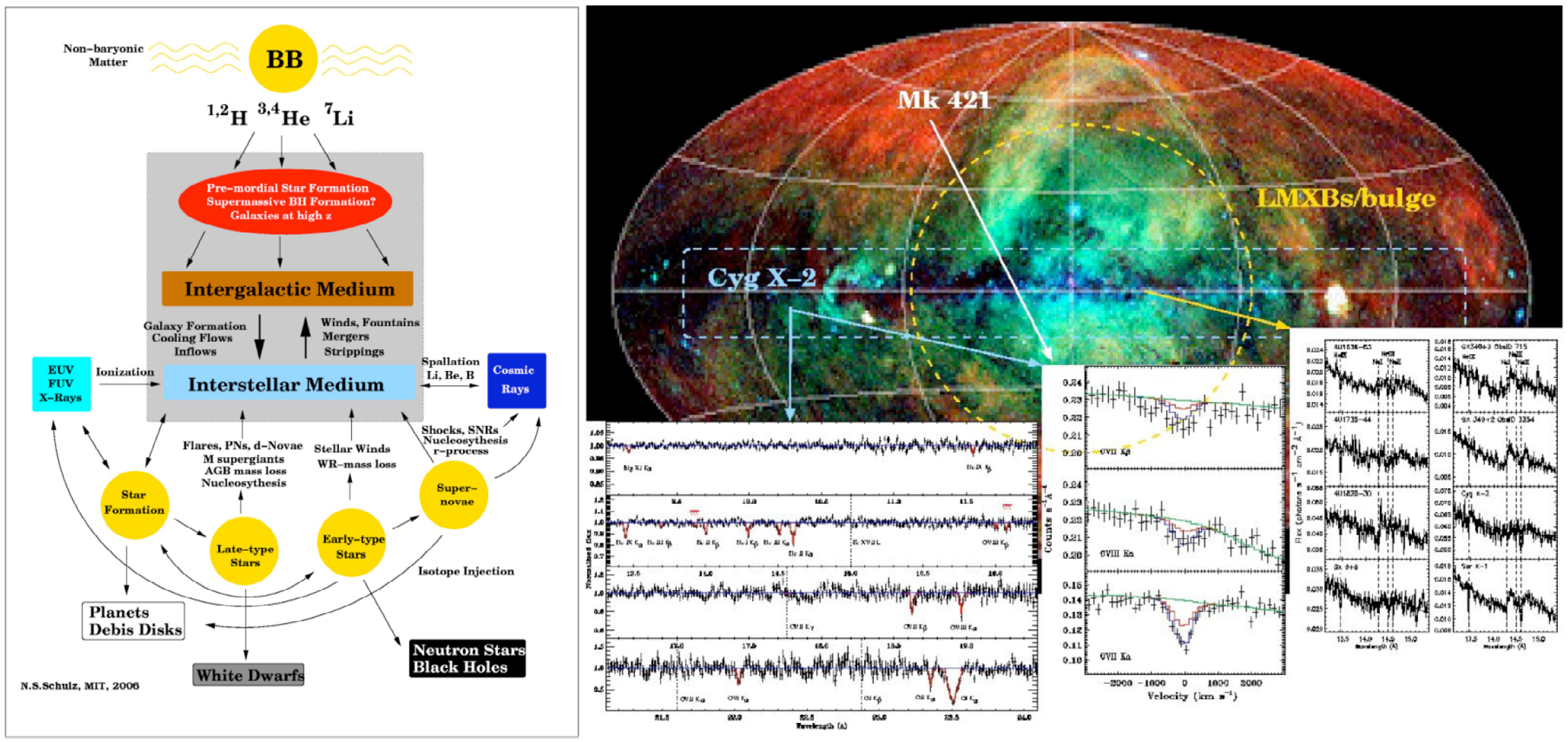

***Figure 1:*** *(left) Schematic diagram illustrating the chemical evolution of ISMs in galaxies and its involved constituents (adapted from Pagel 1991). (right) Rosat All Sky Survey (RASS) image at 3/4 keV with bright X-ray binaries (blue) in the Galactic Halo (yellow circle) and plane (blue square). On the right side a panel of resolved Ne edges and Ne II and III lines from a Chandra survey (Juett et al. 2006). On the left side is a very recent large exposure spectrum of the bright X-ray binary Cyg X-2 (Yao et al. 2009) showing absorption lines from the cool, warm, and hot phase of the interstellar medium. In the middle a comparison of absorption towards Cyg X-2 and Mk 421 indicating that likely all equivalent width is due to galactic absorption.*

A primordial IMF plays an important role in determining the impact of the very first stars on the ionization, thermal, and chemical enrichment history of the intergalactic medium (IGM). It was not until the last decade that it was realized that large fractions of the mass may not be allocated in galaxies but in the IGM, specifically intercluster media (Barkana and Loeb 2001, Madau et al. 2001). There are countless needs and ways where observations contribute to the understanding of primordial chemical evolution, which includes galaxies of all types at varous ranges of redshift and IGMs in the vicinity and line of sight of galaxy clusters, quasars, AGNs, galactic mergers. Standard big bang nucleosynthesis predicts a certain baryon density throughout the evolution of the Universe and observations of H and He absoprtion lines in the Lyman $\alpha$ forest were able to find consistent values at high redshift, i.e above z = 3 (Rauch et al. 1997). However, summing over all the baryons inside stars, neutral atomic gas and molecular gas results in a baryon density that is significantly lower in the local universe. Today there is a growing consensus that these missing baryons reside as part of the IGM in a hot (> $10^6$ K), ionized plasma associated with groups of galaxies (Fang and Canizares 2000) called the warm/hot intergalactic medium (WHIM, IXO white paper by J. N. Bregman et al. 2009a).

The IGM in its past and present is in constant interaction with the interstellar medium (ISM) of galaxies through the process of galaxy formation, cooling flows, and inflows through accretion. On the other hand, ISMs enrich the IGM through fountain flows, winds, mergers and strippings. At the forefront, of course, is the understanding of the ISM of our own Galaxy. In contrast to the IGM, which is almost completely ionized up to redshifts of ~5.5, the ISM is a rather sponge-like structure that contains various phases in terms of structure, temperature, and density. These phases are entangled in various ways with on-going star formation, interactions with stars at almost all evolutionary levels. Critical, but little known are the various ionization fractions in these phases, which should vary based alone on the the fact that the ionization processes likely differ for these phases. Besides a hot phase, which like the IGM is highly ionized, there are warm and cold phases in co-existence which only very lowly ionized and almost neutral matter. Though great strides were made in recent decades to manage to understand some of the complex processes within the local ISM (R < 500 pc, Jenkins & Savage 1974) and some locations in the Galactic plane and halo (Juett et al. 2004, 2006; Yao & Wang 2005), even such low level of detail is still fairly unavailable for the most of the Milky Way and even more for ISMs in other galaxies. With respect to the enrichment of elements through various nucleosynthesis processes it is a daunting fact that we still do not understand why we find so rather different levels of metallicities in galaxies within the Local Universe (observed z < 3).

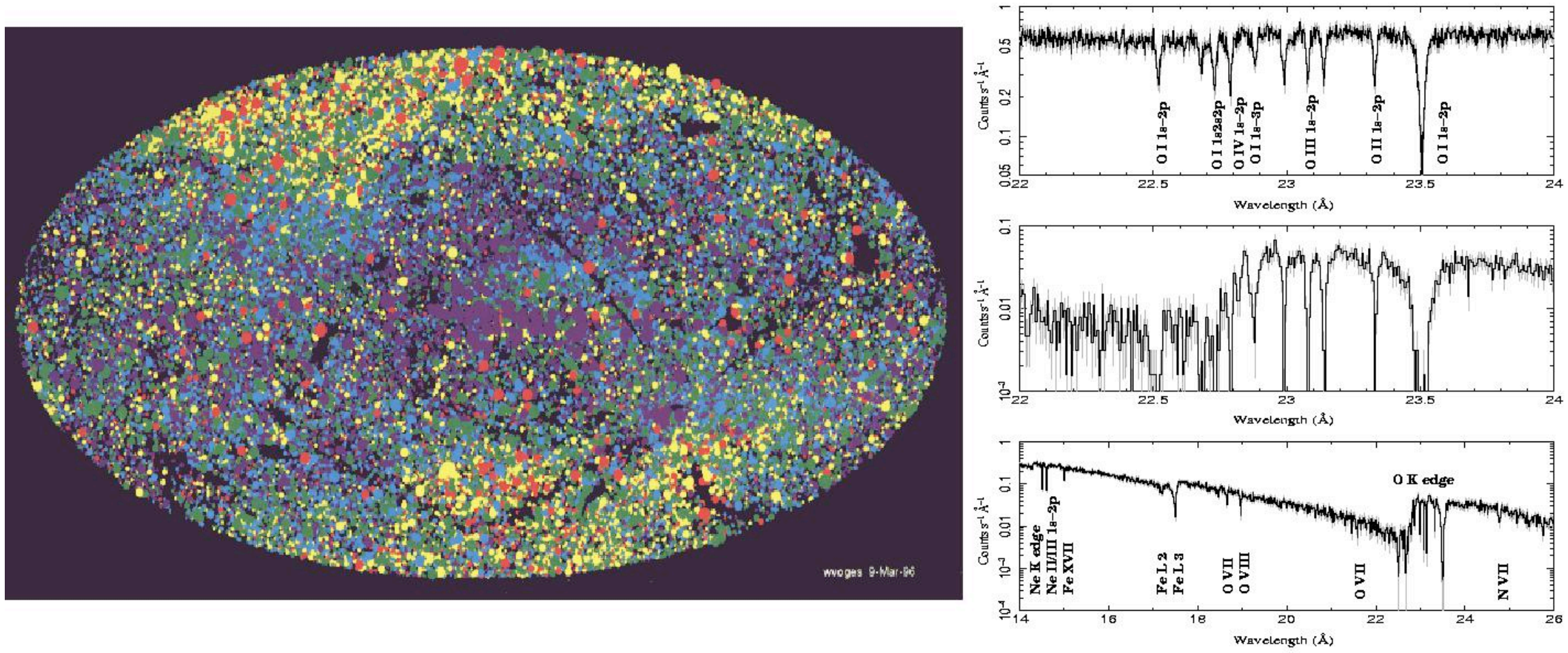

***Figure 2:*** *A plot of over 70,000 sources from the RASS, red sources have low, yellow intermediate, blue have high absorption. (right) Simulated grating spectra from IXO in the range including the Ne K , Fe L, and O K edges (bottom) for a column density of* $N_H = 5x10^{21}$ $cm^{-2}$ *for 30 ks exposure of a* $10^{-10}$ *erg* $cm^{-2}$ $s^{-1}$ *X-ray source in the Galctic plane. The middle panel shows a closeup of the O K edge, the top the O K edge for an absorption of* $N_H = 5x10^{20}$ $cm^{-2}$ *with line identifications.*

ISMs in galaxies contain massive amounts of dust, which manifests itself not only in beautiful optical images of our Milky Way and nearby galaxies. The depeletion of elements in the ISM locked away into dust grain has been a long-standing problem. HST images of the Galaxy's ISM show that depletion levels are different in the hot Galactic halo to the ones in the Galactic disk. While element abundances in the halo show the least deviation to the ones we observe on the atmosphere of our Sun, the ones in the warm and cold disk show depletion levels of various metals of over and factor ten (Frisch 2004). How uniform these depletion levels are nor the nature of the depletion is well known as is the situation outside the local ISM. Specifically in other galaxies where metalicities are know to be different, our knowledge about dust and the depletion of elements needs to be studied (see IXO white paper by J. C. Lee et al, 2009).

## 2. ISM studies in X-rays -- Chandra and IXO Surveys

While EINSTEIN and early rocket flights produced the first X-ray images of supernova remnants (SNRs), missions like ROSAT and ASCA teased us with the possibility to image diffuse matter in X-rays as well as to apply X-ray spectroscopy as a tool to diagnose its properties. The door to study diffuse matter in our Galaxy beyond SNRs and dust scattering in X-rays was pushed wide open by the Chandra X-ray Observatory with its vastly improved X-ray vision and high resolution spectrometers.

There are many ways to study matter in the ISM and IGM in X-rays, which includes emission from shock-heated regions, intermittent absorbers of all temperature ranges, and scattering from immersed dust. All these processes need high levels of throughput and spectral and imaging resolving power. ROSAT (0.1 -- 2.4 keV) allowed first steps in the imaging of hot diffuse matter in galactic halos and hot ionized plasmas in between clusters of galaxies, the generation of the first soft X-ray all sky survey (RASS) that allowed correlations with hydrogen column density surveys (see Figure 1), and first assessments of dust scattering halos (Predehl and Schmitt 1995). The RASS catalog (Voges et al. 1999) contains over 70000 X-ray sources as shown in Figure 2. However, studies of interstellar matter properties are intimately related to spectroscopic resolving powers (R >>1000) which so far has only have barely been achieved with Chandra. The following sections illustrate some recent advances made with Chandra with respect to ISM and IGM spectroscopy and give an the outlook for IXO.

### 2.1 X-ray Surveys of ISMs

The X-ray band is particularly sensitive to K-shell absorption from high abundance elements like C, N, O, Ne, Mg, Si, S, Ar, Ca, Fe, and Ni as well as L-shell absorption of at least Fe. These elements are abundant in the local universe. Resonance absorption and backscattering is also expected from simple and complex molecules that contain one or more of these elements at high densities (see Lee et al. 2009). In order to diagnose ISM properties the technique of backlighting is used, where a strong and distant X-ray source produces a broad continuum which is being absorbed by ISM/IGM elements in the line of sight. A few spectral cases are shown in Figure 1. The left spectral panel shows a recent high resolution X-ray spectrum of Cyg X-2 (Yao et al. 2009, Juett et al. 2004), containing various levels of absorption relating to the cold and warm ISM phases. The right spectral panel in Figure 1 shows several LMXB spectra around the neutral Ne K edge (Juett et al. 2006). These spectra demonstrate the ability to trace all three ISM gas phases in X-ray absorption spectra. In the case of neon, the Ne K edge corresponds to the cold phase, Ne II and III absorption to the warm phase and Ne IX absorption to the hot phase (Juett et al. 2006, Yao and Wang 2005). The scope of these studies is to determine abundances, ionization fractions, turbulence, and dispersion velocities. Typical equivalent widths of the absorption lines range between ~1 and 15 mA and deduced neutral column densities are  of the order of $10^{18}$ $cm^{-2}$ or lower . Ionized lines are up to a factor 25 lower, of the order of $10^{16}$ $cm^{-2}$ . With respect to traditional studies, which used to be performed in the UV for the local ISM (Jenkins & Savage 1974), this method allows long-range scans throughout the entire Galactic plane, which is owed to the fact that X-ray binaries are up to a factor $10^6$ brighter than early type stars and typically many kpc in distance. There are over $2x10^3$ bright sources in our Galaxy that can be efficiently surveyed by IXO.

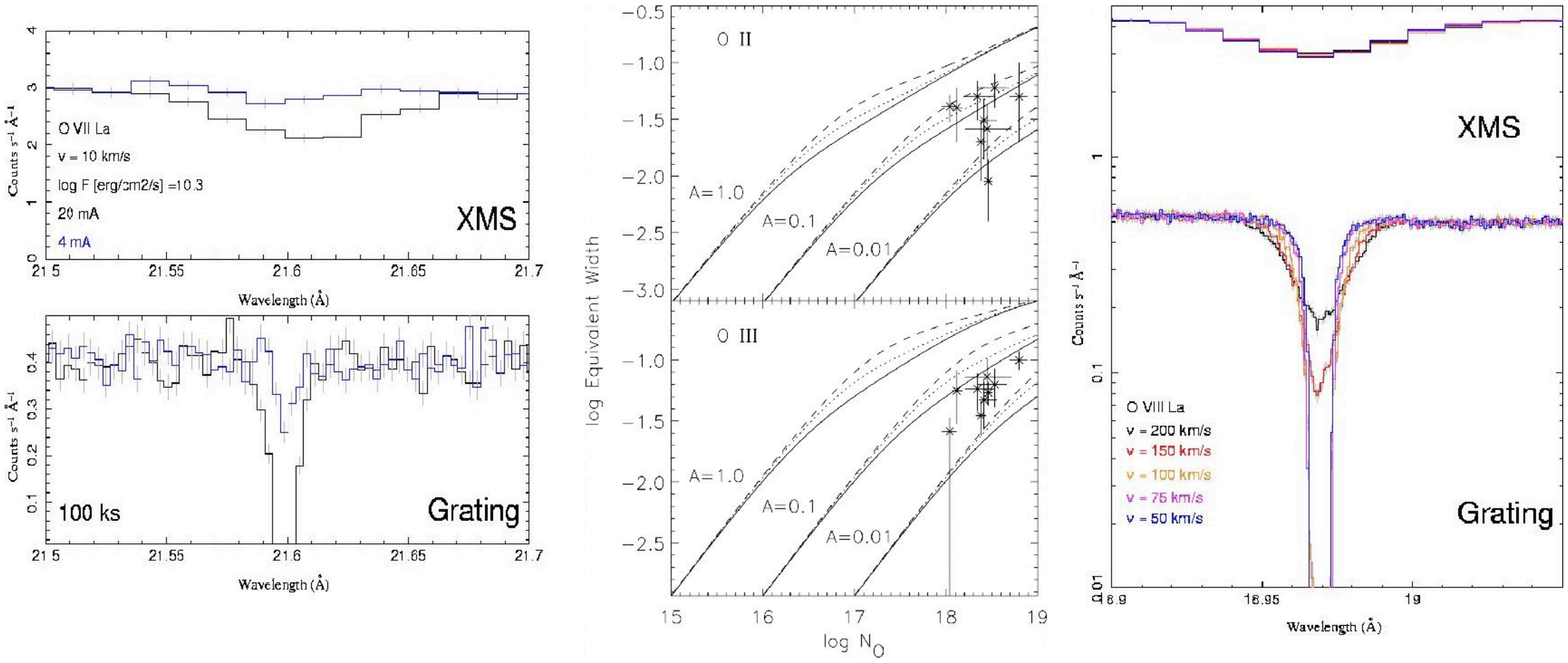

***Figure 3:*** *(right) Simulation of various velocity widths of O VIII absorption. A microcalorimeter is insensitive below 200 km $s^{-1}$, a grating has sensitivity down to 50 km $s^{-1}$. (middle) Curve of growth analysis from Juett et al. (2004) leading to some abundance predictions for O II and III but also indicating the current insensitivity of dispersion velocities between 20 km $s^{-2}$ (solid line) and 200 km $s^{-2}$ (hatched line). (left) Sensitivity comparison of line absorption between a microcalorimeter and a grating. Gratings are sensitive to much smaller line equivalent widths.*

Figure 2 show the distribution of RASS sources with spectral colors. While the white paper by Lee et al. 2009 demonstrates that IXO can determine abundances to precisions of better than 5% by measuring photoelectric edges, the high spectral resolution of the gratings can go much further. Resonance absorption in the cold, warm, and hot phases of the ISM in galaxies as well as the largely ionized IGM are observed from various ions ranging from Carbon to Magnesium and therefore at energies below 1 keV. In order to fully resolve these resonances and measure dispersion velocities of below 200 km $s^{-1}$ a resolution of 3000 is required (see Figure 3). Ionization fractions and dispersion dynamics are important ingredients to gauge star formation rates. Figure 2, 3, 4 summarize the scientific sensitivity of the soft spectroscopic X-ray surveys for our Galaxy and the Local Group, which are aimed to determine:

- fractions and abundance of matter at all ionziation levels in the ISMs in galaxies,
- the distibution of highly ionized matter in extended halos and in between galaxies,
- sizes and scale heights of absorbers ,
- the distribution of dispersion velocities at all ionization levels.

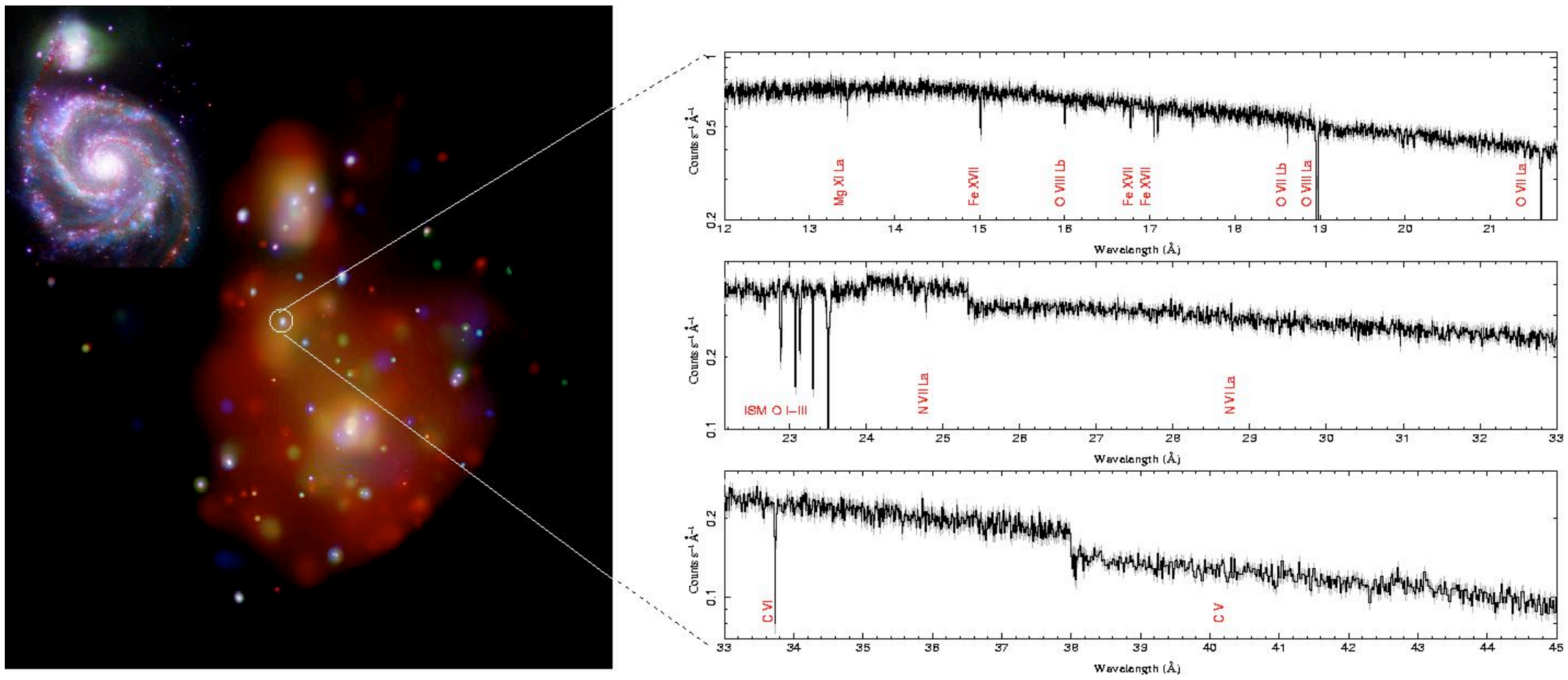

***Figure 4:*** *200 ks simulation of a bright X-ray source in M 51 detecting most expected resonance features.*

These studies are not confined to our own Galaxy and the two Magellanic Clouds but also apply to galaxies in the Local Group. This is illustrated by the simulation in Figure 4. The study of X-ray binary populations using the Chandra and XMM-Newton are currently conducted for many galaxies of the local group. In M33, for example, a total of 260 sources above the flux limit of 3x $10^{-16}$ ergs $cm^{-2}$ $s^{-1}$ have been detected (Grimm et al. 2005) with luminosities larger than log $L_x$ = 34 [erg $s^{-1}$]. M33 and M51 are interesting as here the star formation rates appear significantly higher than in the Galaxy or M31. With IXO we feasibly can survey sources above a luminosity of log $L_x$ = 36 [erg $s^{-1}$] with reasonable exposures (< 600 ks) which applies to roughly $10^3$ X-ray binaries within the local group. Further possibilities to trace cool matter in AGNs has been indicated in MCG6-30-15 in form of the possibility of a properly redshifted Fe L edge (Lee et al. 2001) as well as several cool O absorption lines. In any case, detections of such matter states in AGNs remains an exciting possibility.

### 2.2 X-ray Surveys of the local IGM

Like the Lyman α forest in the optical band, absorbers in the IGM can also produce an X-ray line forest along the line of sight in the X-ray spectrum of a background quasar . Recent observation have claimed such lines for the case of O VIII Lyman α absorption in a Chandra spectrum of the quasar PKS 2155-304 (z~0.06; Nicastro et al. 2002, Fang et al. 2002). Projections from this detection, for example, imply that a large fraction of the missing baryons may be probed by the O VIII absorber. Other detections also include the line of sight towards Mrk 421 (z = 0.03; Williams et al. 2005). While such observations are still rare, they show the potential of X-ray spectroscopy to probe the WHIM (see white paper by Bregman et al. 2009a). What is of interest here is not the WHIM itself but the fact that there will be a manifold of deep observations available to scan the nearby IGM regions through differentiation with local absorptions (see example in Figure 1). Recent detection of O VII and Ne IX Lyman α line towards the X-ray binary LMC X-3 in the Large Magellanic Cloud (Wang et al. 2005) have now spurned interest in such studies. A study by Bregman & Lloyd-Davies (2007) including a number of bright AGNs strongly pointed towards a Galactic Halo origin rather than a diffuse IGM component. Recent results by Yao et al. (2008, 2009) support a local origin. Such absorbers the could be found not only in galactic halos but also in interactive regions of galaxy mergers and strippings and are an exciting aspect of these spectroscopic surveys at very high resolution (see also the IXO white paper by J. N. Bregman et al. 2009b).

### *3. What do we need from Future Observatories*

Future observatories will achieve breakthroughs in X-ray studies of chemical evolution in the Universe once they match the baselines currenly met by the IXO design:

- $A_{eff} > 1000$ cm$^{-2}$ in the wavelength range 10 – 45 Angstrom
  $A_{eff} > 3$ m$^{-2}$ in the wavelength range 1.5 – 10 Angstrom
- R > 3000 in the wavelength range 10 – 45 Angstrom
  R > 1000 in the wavelength range 1.5 – 10 Angstrom
- Calibration requirement ~ 3%

The requirements in the range 10 – 45 Angstrom are most important for elements like C, N, O, Ne, an Mg. The limits on spectral resolution in this range are easiest to understand. The hot phases in the ISM and IGM are turbulent with motions of 100 – 200 km s$^{-1}$, which require R= 3000 and higher to diagnose. With a response accuracy of better than 10% we can see down to ~50 km s$^{-1}$ in Doppler widths (see Figure 3, right). For the cold and warm phase of ISMs this is still challenging as here we expect velocities more of the order of 10 – 60 km s$^{-1}$. Here R > 5000 with an accuracy or 10% is anticipated for Gen-X. At such high resolutions the effective area also becomes critical. An area $A_{eff} > 1000$ cm$^{-2}$ at long wavelengths is generally feasible, but less effective and a goal of 3000 cm$^{-2}$ is anticipated. For the expected line equivalent widths, a 100 ks exposure under the above specification leaves over ~0.05 cps within one R=3000 resolution bin for a typical X-ray source at 10 Mpc distance. This limits feasible studies to the Local Universe.